\documentclass[aps,prl,twocolumn,preprintnumbers,floatfix,reprint]{revtex4-1}
\usepackage{mathrsfs,natbib}
\usepackage{amsmath,amssymb, amsfonts, bbm}
\usepackage{graphics,epsfig,color,subfigure}
\usepackage{hyperref}
\usepackage{verbatim}
\usepackage{eufrak}

\newcommand{\be}{\begin{equation}}
\newcommand{\ee}{\end{equation}}

\def\bea{\begin{align}}
\def\ena{\end{align}}

\def\beqa{\begin{eqnarray}}
\def\enqa{\end{eqnarray}}

\def\Z{\mathbb{Z}} 
\def\R{\mathbb{R}}

\def\F{{\cal{F}}}

\def\Z{\mathbb{Z}} 
\def\R{\mathbb{R}} 
 
\def\B{\mathbb{B}}

\def\Im{\text{Im}}
\def\Re{\text{Re}}
\def\tr{\text{tr}}

\def\bar{\overline}

\def\th{{\theta}}

\def\l{{\lambda}}

\hypersetup{
    colorlinks=true,       
    linkcolor=red,          
    citecolor=blue,        
    filecolor=magenta,      
    urlcolor=blue           
}

\begin{document}

\title{Resurgence in QFT: Unitons, Fractons and Renormalons in the Principal Chiral Model}

\author{Aleksey Cherman$^1$}
\email{cherman@physics.umn.edu} 
\author{Daniele Dorigoni$^2$}
\email{d.dorigoni@damtp.cam.ac.uk} 
\author{Gerald V. Dunne$^3$}
\email{dunne@phys.uconn.edu}
\author{Mithat \"{U}nsal$^4$}
\email{unsal.mithat@gmail.com} 
\affiliation{
$^1$ Fine Theoretical Physics Institute, Department of Physics, University of Minnesota, USA\\
$^2$ DAMTP, University of Cambridge, Wilberforce Road, Cambridge CB3 0WA, UK \\
$^3$ Physics Department, University of Connecticut, Storrs, CT, 06269 USA\\
$^4$ Department of Physics and Astronomy, SFSU, San Francisco, CA 94132 USA}

\preprint{FTPI-MINN-13/28, UMN-TH-3218/23}
\preprint{DAMTP-2013-40 }

\begin{abstract}
We explain the physical role of non-perturbative saddle points of path integrals in theories without instantons, using the example of the asymptotically free two-dimensional principal chiral model (PCM). Standard topological arguments based on homotopy considerations suggest no role for non-perturbative saddles in such theories. However, resurgence theory, which unifies perturbative and non-perturbative physics, predicts the existence of several types of non-perturbative saddles associated with features of the large-order structure of perturbation theory. These points are illustrated in the PCM, where we find new non-perturbative `fracton' saddle point field configurations, 
and give a quantum interpretation of previously discovered `uniton' unstable classical solutions.  The fractons lead to a semi-classical realization of IR renormalons in the circle-compactified theory, and yield the microscopic mechanism of the mass gap of the PCM. 
\end{abstract}

\maketitle

{\bf Introduction:} 
In general, observables in quantum field theories (QFTs) receive perturbative and non-perturbative contributions. The perturbative contributions summarize information about quantum fluctuations around the trivial perturbative saddle point (vacuum) of the path integral, while the non-perturbative contributions come from quantum fluctuations around the non-trivial non-perturbative (NP) saddle points.  In this paper we develop a deeper understanding of what the structure of perturbation theory implies for the nature and existence of NP saddle points of path integrals.

We illustrate these ideas with the two-dimensional (2d) $SU(N)$ principal chiral model  (PCM).  The PCM is an asymptotically free matrix field theory, and is believed  to have a dynamically generated mass gap determined by the strong scale $\Lambda= \mu e^{- \frac{4 \pi}{g^2(\mu) N}}$, with $\mu$ the renormalization scale, see e.g. \cite{Faddeev:1985qu,Fateev:1994ai,*Fateev:1994dp,Orland:2011rd,*Cubero:2012xi,*Cubero:2013iga}. The PCM models many features of  4d Yang-Mills (YM)  theory,
but historically it has received less attention \footnote{Except for its deformation by a Wess-Zumino-Novikov-Witten term, see e.g. \cite{Tsvelik:1996zj}.} 
than its vector-model cousin, the ${\mathbb C \mathbb P^{N-1}}$ model,  because:
{\it (i)} since $\pi_2[SU(N)] = 0$, there are no topologically stable instanton configurations which may lead to NP factors such as  $e^{- t/g^2}$;
  {\it (ii)} its large-$N$ limit is not analytically tractable \cite{Polyakov:1987ez}.  The ${\mathbb C \mathbb P^{N-1}}$ model has neither of these issues, but {\it(i)} is also shared with 
many other  2d QFTs, such as $O(N>3)$ and $Sp(N)$ models, which are relevant to condensed matter physics\cite{Tsvelik:1996zj}.

However, the divergent structure of perturbation theory in the PCM is very similar to YM or ${\mathbb C \mathbb P^{N-1}}$.  After regularization and renormalization, the perturbative series has at  least two types of factorial divergences.
One is due to the 
 combinatorics of the Feynman diagrams, while the other is known as  the  IR and UV renormalon divergence \cite{tHooft:1977am,Fateev:1994ai, Fateev:1994dp} and comes from the low and high momenta in phase space integrals.  Resummation of the perturbative series using the standard technique of Borel summation leads to singularities in the Borel plane. Infrared (IR) renormalons render the Borel sum ambiguous and ill-defined, because there is a subset of factorially divergent terms that do not alternate in sign.
 These problems are ubiquitous in asymptotically free QFTs, including YM and QCD \cite{Marino:2012zq}, as well as in string theory \cite{Aniceto:2011nu}.

It is generally believed that in quantum mechanics (QM) and QFT, the ambiguities in the summation of perturbative series due to the growth in the number of Feynman diagrams cancel against ambiguities in the contributions from NP instanton-anti-instanton saddle points,   $[\cal I  \overline {\cal I}]$, $a_n \sim n! / (S_{[\cal I  \overline {\cal I}]})^n$~\cite{tHooft:1977am}. On the other hand, the semiclassical meaning of IR renormalons 
has been unclear until recently, when it was  shown that  renormalons may also be
 continuously connected  to  new semi-classical  NP saddle points \cite{Argyres:2012vv,Argyres:2012ka,Dunne:2012ae, Dunne:2012zk}.  In the weak coupling regime of circle compactified   deformed YM and QCD(adj) in 4d, and 
 the  ${\mathbb C \mathbb P^{N-1}}$ model  in 2d,  it was shown that   BPST  instantons fractionalize into  $N$  monopole-instantons ${\cal M}_i$ 
  \cite{Lee:1997vp, Kraan:1998pm},    and    $N$ kink-instantons ${\cal K}_i$ \cite{Bruckmann:2007zh, Brendel:2009mp, Harland:2009mf, Dabrowski:2013kba},  respectively. 
  Correlated $[{\cal K}_i \overline {\cal K}_i ] $ and $[{\cal M}_i \overline {\cal M}_i ] $ events  control the  IR-renormalon singularities in these theories, and render physical observables unambiguous through the mechanism of resurgence \cite{Ecalle:1981,Dunne:2012ae, Argyres:2012ka}.

However,  the PCM has neither instantons nor fractional instantons. 
 In fact,  the PCM has no known stable NP saddles  which could lead to NP factors such as  $e^{- t/g^2}$.   This produces a deep puzzle. Since perturbation theory is divergent and non-Borel summable,  an attempt to do Borel resummation results in ambiguities of the form $\pm i e^{-t_i/g^2}$.  
   If the theory is to be semi-classically meaningful and well-defined according to the criterion of \cite{Dunne:2012ae, Argyres:2012ka},   such NP ambiguities   {\it must} cancel, i.e.,    there must exist NP saddles whose amplitude is proportional to     $\pm i e^{-t_i/g^2}$.  This is a  highly non-trivial prediction of  resurgence theory applied to QFT. But since there are no instantons, what are these NP saddles?

Thus the perturbative similarity (in particular the non-Borel-summability due to IR renormalons) between 
the PCM and other asymptotically free theories such as YM and ${\mathbb C \mathbb P^{N-1}}$, appears to be in conflict with their NP difference: YM and ${\mathbb C \mathbb P^{N-1}}$ have non-trivial homotopy groups, and hence instantons, while the PCM has trivial homotopy, $\pi_2(SU(N))=0$, and no instantons. This suggests that topology alone is insufficient to fully characterize NP saddles, and misses a large class of important NP saddle points. 
In this work, we combine resurgence theory with a physical principle of continuity, and show the existence of new  NP   saddles in the path integral of the PCM, which we refer to as `fractons' following the groundbreaking early work \cite{Shifman:1994ce}; see also \cite{Smilga:1994hc}. 
Our analysis easily generalizes to other theories,  such as $O(N>3)$ or $Sp(N)$ models, which also have no instantons.

{\bf Unitons:} 
The PCM has classical action
\begin{align}
\label{eq:BosonicSigmaModel}
S_b = \frac{N}{2 \lambda} \int_{M} d^2x 
 \, \tr \,\partial_{\mu} U \partial^{\mu} U^{\dag}, \;\;  U \in SU(N), 
 \end{align}
where  $\lambda = g^2 N$ is a dimensionless coupling constant, and we work in Euclidean space with $M = \mathbb{R}^2\textrm{ and } \mathbb{R} \times S^1$. The PCM has the symmetry $SU(N)_L \times SU(N)_R$ acting as $U \to g_L U g_R^{\dag}$.
There are no  instantons, but there exist `unitons', 
{\it finite-action} solutions to the {\it second order} Euclidean equations of motion, discovered in the seminal work of Uhlenbeck \cite{uhlenbeck1989harmonic}. Further properties are discussed in
\cite{Valli1988129,Piette:1987qp,Piette:1987qr,Wood01051989,Ward:1990vc,Dunne:1994uy,Dunne:1995ai}. 
These uniton solutions, which are harmonic maps from $S^2$ into $SU(N)$,  did not receive much attention in the QFT literature,  mainly because 
they are unstable; small fluctuations lead to a decrease in the action \cite{Piette:1987ia}.
Topologically, this  instability is expected, since instantons would be smooth maps from compactified base space $M=\mathbb{R}^2 \cup \infty = S^2$ to the target space $\mathcal{T} = SU(N)$, and classified by the second homotopy group $\pi_2[SU(N)]$.  But  $\pi_2[SU(N)] = 0$, so there are no topologically stable instantons.     

However, the notion that all finite action saddles  contributing to path integrals are classified by $\pi_d[\mathcal{T}]$ is incorrect, even if $\pi_d[\mathcal{T}]$ is non-trivial, as pointed out in \cite{Dabrowski:2013kba} for the $\mathbb{CP}^{N-1}$ model, which has exact finite action non-BPS solutions, and where the connection to resurgence was also emphasized.  In fact, these points can already be seen in quantum mechanics (QM). Consider an instanton in  QM with a periodic potential and coupling $g$, where instantons are classified by their winding number $W \in \pi_1(S^1) = \mathbb{Z}$, and the basic instanton solution has $W=1$. This is a solution to the first order BPS equation, and possesses an exact zero mode, its position.  The amplitude for this event is ${\cal I} \sim e^{-S_{\cal I} + i \Theta}$, where $\Theta$ is the topological $\Theta$ angle \cite{Tsvelik:1996zj}.  Now, consider a correlated 
 instanton-anti-instanton event $[{\cal I} \overline {\cal I}]$.   
  This is {\it topologically indistinguishable} from the perturbative vacuum, since $W=1+ (-1) = 0$, but its action is $S = 1+1=2$, in units of the instanton action.  
 Yet the separation between the two instantons is a negative quasi-zero mode,  and the action of this saddle decreases with decreasing separation.
   To write the two-event amplitude one must integrate over the quasi-zero mode.   Naively,  when $\arg(g^2) =0$ this integration is dominated by short distances, and is 
   ill-defined.
   However, doing the quasi-zero mode integration at  $\arg(g^2) = 0^{\pm}$, we find $[{\cal I} \overline {\cal I}]_{\pm}  \sim \left( \log \frac{1}{g^2} - \gamma \pm  i \pi  \right) e^{-2 S_{\cal I}}$, a two-fold ambiguous result. This is a manifestation of the fact that 
 $\arg(g^2) = 0$ is a Stokes line.  Resurgence theory explains that the purely imaginary ambiguous part  of the non-perturbative amplitude cures  the ambiguity associated with the 
non-Borel summability of  perturbation theory,   i.e.,  $\Im  \left(  \B_{0, g^2=0^\pm}+ [{\cal I} \overline {\cal I}]_{\pm}   \B_{2, g^2=0^\pm}   \right) =0$, where  $\B_{0, g^2=0^\pm}$ and  $\B_{2, g^2=0^\pm}$  are left/right Borel sums of the formal perturbative series describing quantum fluctuations  around the perturbative and non-perturbative 
 $[{\cal I} \overline {\cal I}]_{\pm}$ saddle points of the path integral, respectively \footnote{The real part of $\B_{2, g^2=0^\pm}$  contributes an exponentially small NP term, 
$e^{-2 S_{\cal I}}$ to physical observables}.  For a fuller discussion of this cancellation mechanism, see   \cite{Dunne:2012ae}. Thus, the `instability' of the $[{\cal I} \overline {\cal I}]$ saddle point, {\it i.e.}, a negative mode in the fluctuation operator, 
is in fact a positive feature, not a deficiency.  Without it, the theory would be ill-defined.    
 
 \begin{figure}[htbp]
\includegraphics[width=0.23\textwidth]{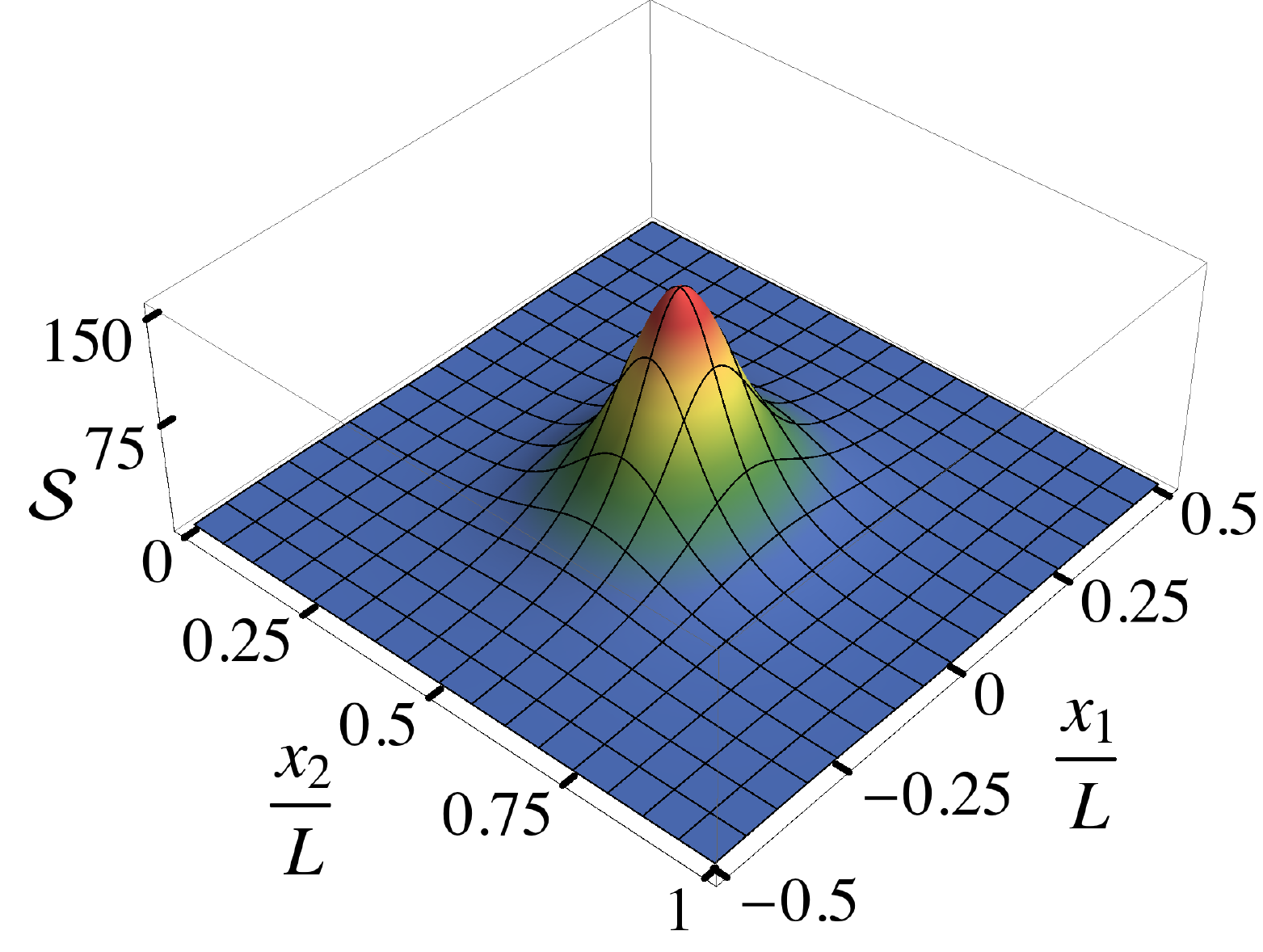}
\includegraphics[width=0.23\textwidth]{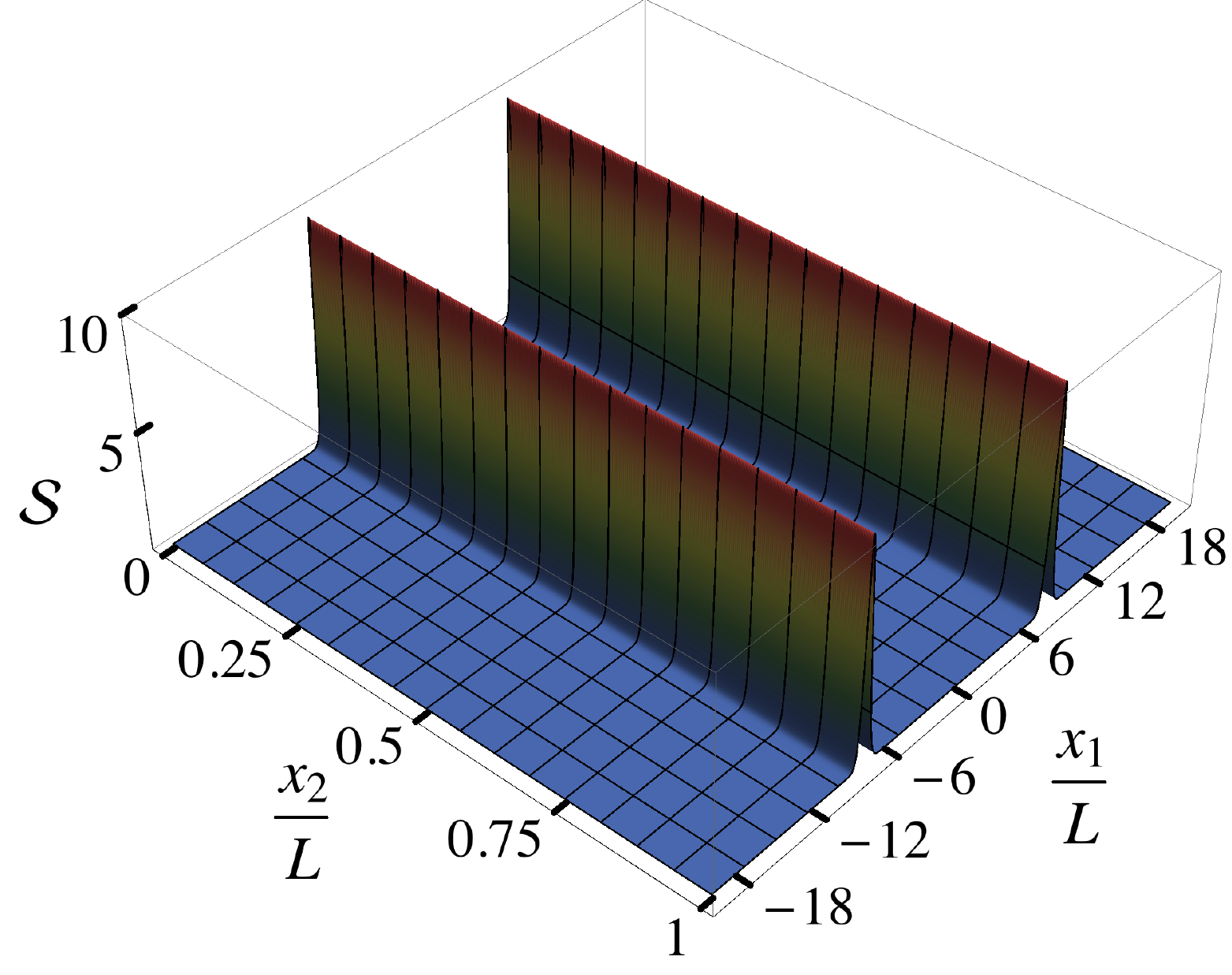}
\caption{Action densities $\mathcal{S}$ for small (left) and large (right) $SU(2)$ unitons in the setting described in the text. The large uniton splits into two fractons. 
}
\label{fig:SU2Uniton}
\end{figure}

 \begin{figure}[htbp]
\includegraphics[width=0.23\textwidth]{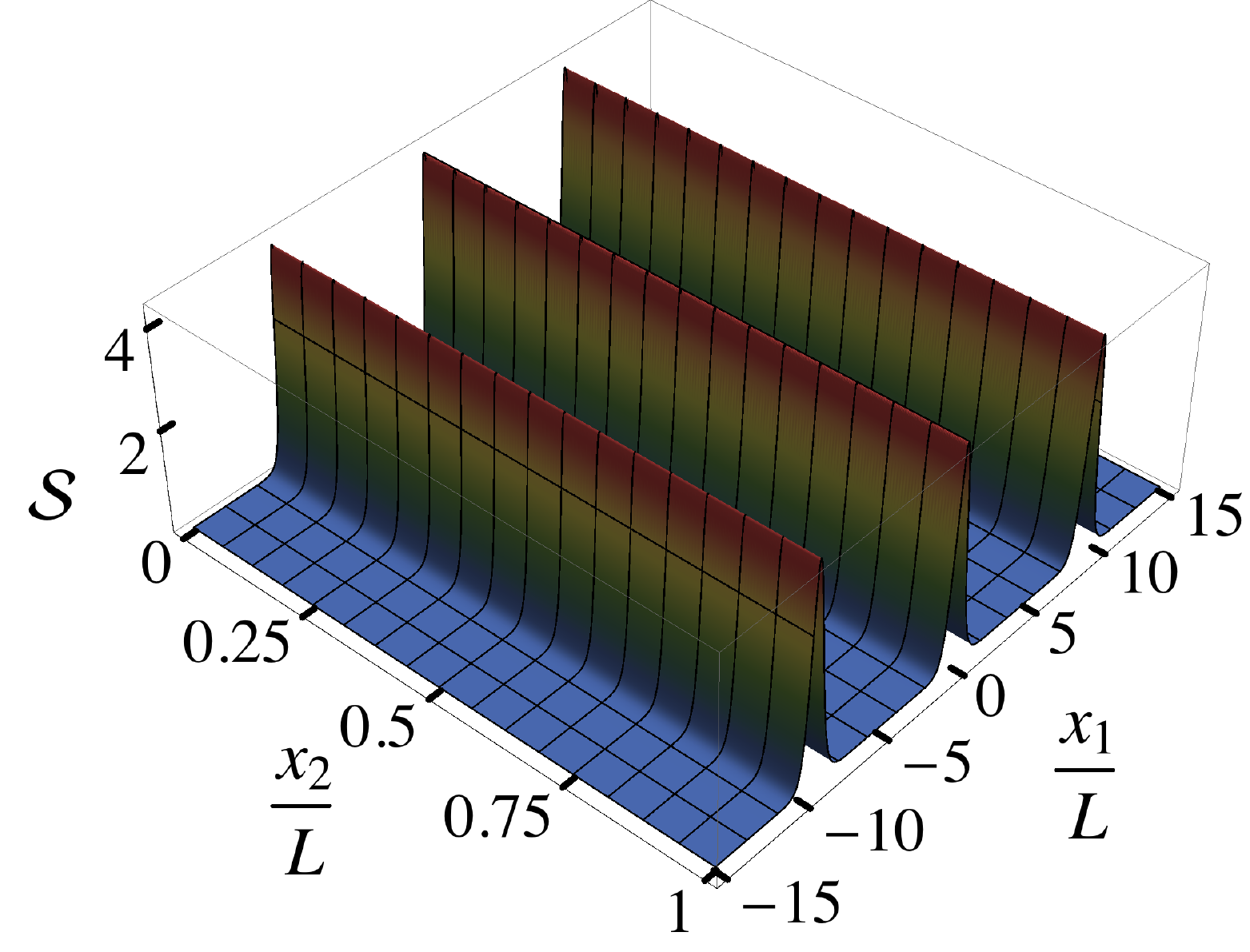}
\includegraphics[width=0.20\textwidth]{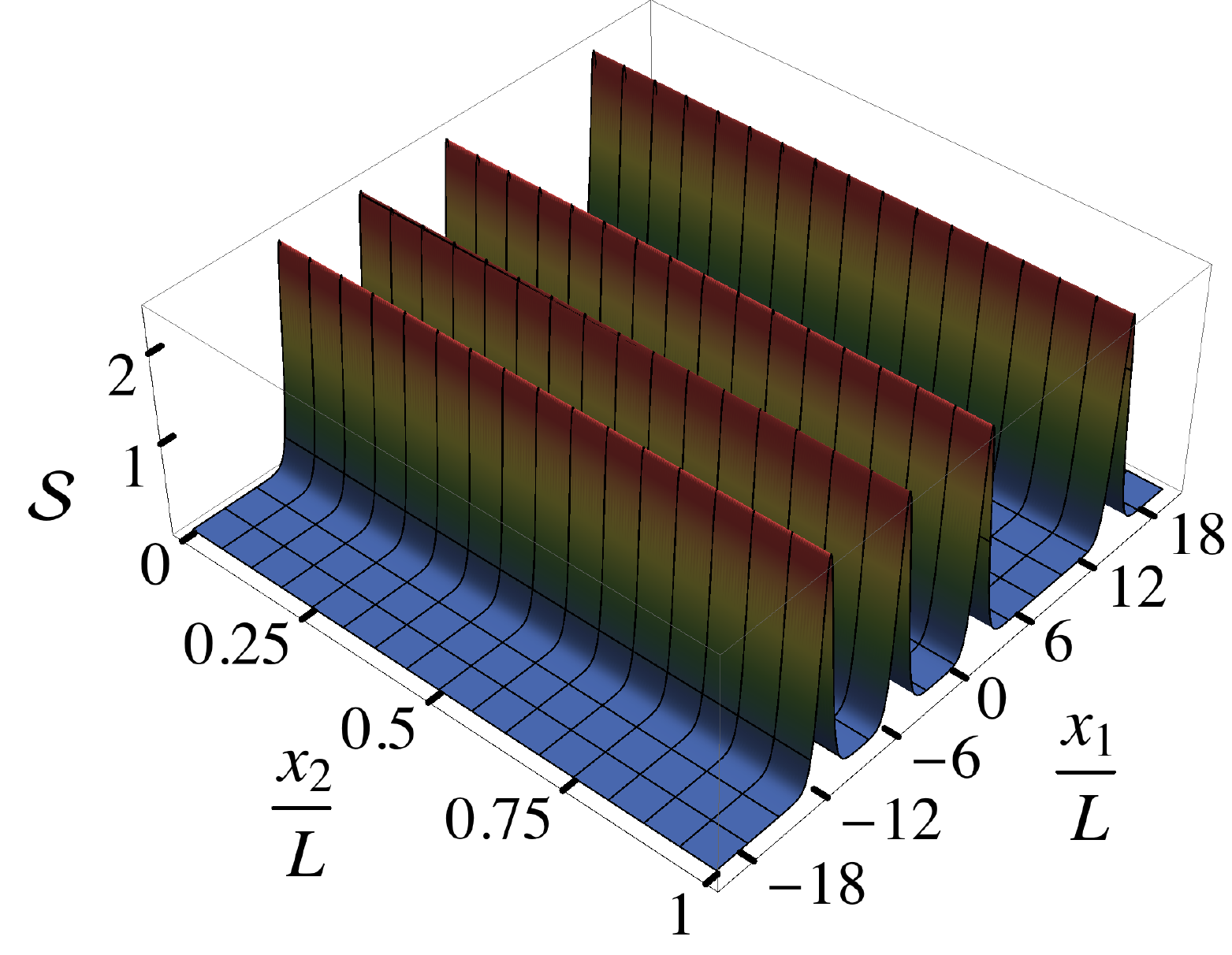}
\caption{Action densities $\mathcal{S}$ for large $SU(3)$ and $SU(4)$ unitons, which split into three and four fractons respectively. 
}
\label{fig:SU3Uniton}
\end{figure}

 Unitons, as finite action non-BPS  field configurations (just like $[{\cal I} \overline {\cal I}]$ events), must be summed over in  a semiclassical analysis of the path integral.  The uniton action is quantized in units of $S_{\mathcal{U}} \equiv \frac{8 \pi}{g^2}$ \cite{Valli1988129}.
The minimal  uniton solution is easy to obtain. Let $v(z) \in \mathbb{C}^N$, with $z = x_1 + i x_2$, $x_{\mu} \in M$, be a single instanton solution in  ${\mathbb C \mathbb P^{N-1}}$~\cite{DAdda:1978uc}.  
Then the minimal uniton in the $SU(N)$  PCM is  given by  $U(z, \bar z) = e^{i\pi/N} (1-2\mathbb{P})$, where $\mathbb{P}$ is the projector: $\mathbb{P}_{ij}= \frac{v_i v^{\dagger}_j} {v^{\dagger}\cdot v}$. 
Fig.~\ref{fig:SU2Uniton} (left) depicts a small uniton in $SU(2)$.  

The uniton amplitude provides  a  substitute for  instantons in theories with a trivial homotopy group, as will be demonstrated below. 
  We write the amplitude associated with the basic uniton event and observe its relation to the strong scale  
  \begin{align} 
   {\cal U}  (\mu) \sim   \det [ O_{\cal U} (\mu) ]  \;  e^{- \frac{8 \pi}{g^2 (\mu)}}, \qquad  \Lambda^{2 \beta_0} = \mu^{ 2 \beta_0}     {\cal U} \quad ,
   \end{align}
where  $\det O_{\cal U}$ encodes the Gaussian fluctuations around the uniton saddle point.  In contrast, for theories with instantons, 
   we would have  $\Lambda^{ \beta_0} = \mu^{  \beta_0}     {\cal I} $, and $\beta_0 = N$ is the one-loop $\beta$-function of the theory.    
   
{\bf Fractons.} 
We cannot directly study the dynamics of the theory on $\mathbb{R}^2$, because the PCM becomes strongly coupled at large distances, just like YM and ${\mathbb C \mathbb P^{N-1}}$. 
However,  there exists a way to continuously connect the strongly coupled PCM on $\mathbb{R}^2$ to a weakly-coupled calculable regime,  
analogous  to the double-trace deformation of YM  theory \cite{Unsal:2008ch}.  
Then, we can address many NP questions in weak coupling, and by adiabatic continuity, extract universal 
results valid even at strong coupling.  This set-up also permits us to see the fractionalizaton of unitons into their constituents, the \emph{fractons}.  The construction involves   introducing the twisted boundary conditions    
$U(x_1, x_2+L) = e^{i H}  U(x_1, x_2) e^{-i H} $, 
 $e^{i H} =  {\rm Diag} \Big[e^{2\pi i\mu_1}, e^{2\pi i\mu_2}, \ldots, e^{2\pi i\mu_N} \Big]$, or equivalently turning on an  $x_2$ component of a background gauge field for the $SU(N)_V$ symmetry so that $\partial_\mu U \rightarrow  D_{\mu}U = \partial_\mu  U + i  \delta_{\mu 2} L^{-1} [H, U]$.  Then we study  the dynamics of the action $S_b = \frac{N}{2 \lambda} \int_{M} d^2x  \, \tr |D_{\mu} U|^2$.  It turns out that there is a unique choice for the $\mu_i$, determined by the condition of  unbroken $\Z_N$ center symmetry, such that the small-L theory is continuously connected to its large $L$ limit, without phase transitions or rapid cross-overs as a function of $L$.  For more details see \cite{CDUlong}.

Working with the special small-L theory defined above, the easiest way to show the splitting  of a uniton into fractons is the following. Let $v_{\rm tw}(z)$ be a single instanton solution in  ${\mathbb C \mathbb P^{N-1}}$ on  $\mathbb R \times S^1$ in the presence of twisted boundary conditions,  which exhibits fractionalization of an instanton into kink-instantons as the size moduli of the instanton is changed from small to large \cite{Bruckmann:2007zh,Brendel:2009mp,Harland:2009mf}, \cite{Dunne:2012zk}.  
   Then, the uniton in the $SU(N)$  PCM is  given by  $U_{\rm tw}(z, \bar z) = e^{i \pi/N} (1-2\mathbb P_{\rm tw})$, where $\mathbb P_{\rm tw}= \frac{v_{\rm tw} (v_{\rm tw})^{\dagger}} {(v_{\rm tw})^{\dagger} v_{\rm tw}}$.   Fig.~\ref{fig:SU2Uniton} (right)  and  Fig.~\ref{fig:SU3Uniton} depict the fracton constituents of a uniton for  $N=2, 3,4$. 

It is straightforward to construct explicit solutions corresponding to isolated fractons in the $SU(N)$ PCM.  For instance, for $SU(2)$, using Hopf coordinates $\theta, \phi_1, \phi_2$ the action is 
\begin{align}
 S=   \frac{1}{ g^2} \int_M \left[ (\partial_\mu \theta)^2 + \cos^2 \theta   (\partial_\mu \phi_1)^2 +  
\sin^2 \theta   (\partial_\mu \phi_2 + \xi \delta_{\mu2})^2 \right]  \qquad
\nonumber
\end{align} 
where $  \xi =  2 \pi (\mu_{2} - \mu_1) L^{-1}$. 
In the small-$L$ regime, forgetting about the high Kaluza-Klein modes,  we land on QM 
with a   non-trivial  potential on the $SU(2)$ manifold: 
$
S= \frac{L}{g^2}\int_{\R }  \Big[  \dot \theta ^2 +   \cos^2 \theta \dot \phi_1^2  +    \sin^2  \theta    \dot \phi_2^2
+ \xi^2 \sin^2 \theta     \Big] ,   $
where the crucial existence of the potential term is due to the non-trivial background holonomy. 
The equations of motion associated with this action admit the solution  $\phi_{1,2}=\phi_{1,2}^{0} $ 
and  $\theta(x_1; x^{(0)}_1) = 2 {\rm arccot} \left[e^{-\xi (x_1 - x_1^{(0)})} \right] $. The constants of integrations 
$\{ \phi_{1}^{0}, \phi_{2}^{0},  x^{(0)}_1\} $ are  the  three zero modes associated with a fracton.

As in gauge theory on $\R^3 \times S^1$ and the  ${\mathbb C \mathbb P^{N-1}}$ model on  $\R \times S^1$, where there exist Kaluza-Klein (KK) monopole-instantons\cite{Lee:1997vp,Kraan:1998pm} and KK kink-instantons\cite{Bruckmann:2007zh, Brendel:2009mp,Harland:2009mf} respectively, which are associated with the affine root of the $SU(N)$ algebra, there is also a KK-fracton in the PCM.   Taking this into account, there are $N$ basic types of fractons in the $SU(N)$ PCM in a ${\mathbb Z_N}$ symmetric background, each of which carries $\frac{1}{N}$ of the action of a uniton.
Namely, 
\begin{equation}
{\cal F}_i \sim   e^{- \frac{8\pi(\mu_{i+1}- \mu_i)}{g^2} } \sim   e^{- \frac{8\pi}{g^2N} } , \qquad {\cal U}=  \prod_{i=1}^{N} {\cal F}_i 
\label{fracton}
\end{equation}
The surprise here with respect to earlier work  \cite{Lee:1997vp,Kraan:1998pm, Bruckmann:2007zh, Brendel:2009mp,Harland:2009mf,Dabrowski:2013kba,Gorsky:2013doa} is that we are now considering a theory which does not have instantons.    Since each fracton carries three zero modes, and each uniton is composed from $N$ fractons, the number of the combined  zero and quasi-zero mode of a uniton must be $3N$. This is analogous to what happens in the ${\mathbb C \mathbb P^{N-1}}$ model, where each instanton has $2N$ exact zero modes, and each kink-instanton has two  zero modes.

{\bf Renormalon  and uniton ambiguities on $\R^2$:}  The IR renormalon divergence and ambiguities  on $\R^2$ can be determined in two different ways.   One is by studying a sub-class of planar Feynman diagrams.  The number of planar diagrams grows only exponentially \cite{Koplik:1977pf,Brezin:1977sv},  but a subset of such diagrams contribute factorially due to momentum integration at large orders, hence the effect is present at large-$N$ as well \cite{Marino:2012zq}. 
Another way is  using  the S-matrix and Bethe-Ansatz equations (starting with the standard  assumptions thereof, such as that
 the theory is gapped).  The approaches must give the same answer, but for the PCM the second approach has been the main one pursued, with the result that the non-alternating late terms in perturbation theory diverge as 
  $   n!  (\frac{\lambda}{8 \pi})^n  $,  meaning that perturbation theory is non-Borel resummable \cite{Fateev:1994dp,Fateev:1994ai}.
This produces an  ambiguity  of the form   $\pm i  e^{ - \frac{ 8\pi k }{g^2 N}}$.  The IR renormalon singularities found in \cite{Fateev:1994dp,Fateev:1994ai} lie on the positive real Borel axis at
\begin{align}
   \label{sing-1}
\mathbb{R}^2: \;\;  t_k^{+} =  8 \pi k/N  = k  [g^2 S_{\cal U}]/\beta_0,     \qquad k \in \mathbb Z^{+}   
\end{align}
 The appearance of the 't Hooft coupling in the IR renormalon ambiguity  means that, unlike instanton-anti-instanton and  uniton ambiguities, 
 it   does not go away in the large-$N$ limit.  
 Also note that  the  leading IR-renormalon singularity is proportional to the square of the strong scale, $ e^{ -  \frac{ 8\pi  }{g^2 (Q) N} }  \sim   (\Lambda/Q)^2$ where $Q$ is the Euclidean momentum. 
 
  In   theories with instantons and a non-trivial homotopy group $\pi_d$,   the leading  IR-renormalon ambiguity   is approximately 
  $e^{-S_{{\cal I}\overline {\cal I} }/\beta_0} = e^{-2S_{\cal I} / \beta_0 }$ 
 and is exponentially larger than the  $ [{\cal I}\overline {\cal I}]_{\pm}$ ambiguity, as emphasized by 't Hooft \cite{tHooft:1977am}. 
   In the PCM, the relation between the leading renormalon and uniton ambiguity  is   $e^{ -  \frac{ 8\pi  }{g^2 N} } \sim e^{-S_{\cal U}/\beta_0}$. 
 Since $\pi_2 [\mathcal{T}]$ is trivial for the PCM, 
 there is nothing preventing a uniton from appearing as a singularity in the Borel plane associated with the perturbative sector. This is to be contrasted with instantons, which carry a topological charge, and cannot appear as a singularity in the Borel plane.  Indeed, on $\R^2$, we expect a pole associated with  $[\cal U]_{\pm}$ upon integration over quasi-zero modes, related to the combinatorics of Feynman diagrams.  The cancellation mechanism for the IR renormalon ambiguities on  $\R^2$ is unknown, but after spatial compactification to  $\R \times S^1$ the theory is under control.  Below, we provide a microscopic mechanism of cancellation   on $\R \times S^1$ in the regime of the theory  continuously connected to $\R^2$.

 
 {\bf Continuity and cancellation of  semi-classical renormalon ambiguities on $\R \times S^1$:}   At small-$L$, the theory reduces to QM, which is continuously connected to the 2d QFT. Consider the ground state energy $\mathcal{E}$. The late terms of perturbation theory for $\mathcal{E}$
involve a non-alternating divergent subseries. Upon left/right Borel resummation ${\cal S}_{0^\pm }  {\cal E}$ at $ g^2 + i 0^{\pm}$, we find a two-fold ambiguous result \cite{CDUlong}:
  \begin{align} 
{\cal S}_{0^\pm }  {\cal E}  (g^2) 
= \Re\,  \B_0   \mp  i  \frac{32 \pi}{  g^2N } e^{-\frac{16 \pi}{g^2N}} 
\label{eq:s1}
\end{align}
reflecting the non-Borel summability of the theory on the $\arg(g^2)=0$ Stokes line.   This is the semi-classical realization of the renormalon ambiguity. The  associated semi-classical singularities in the Borel plane are located at 
 \begin{align}
\mathbb{R} \times S^1: \;  t_k^{+, \rm s.c.} =  \frac{16 \pi k}{N}  = 2 \times k \times  \frac{g^2 S_{\cal U}}{\beta_0},    \; k \in \mathbb Z^{+}  
   \end{align}
diluted by a factor of two with respect to $\R^2$, but  parametrically in the same neighborhood as the IR renormalon singularities of 't Hooft seen in (\ref{sing-1}). 

Remarkably,  as predicted by the resurgence theory of  \'Ecalle\cite{Ecalle:1981}, this ambiguity cancels against the fracton-anti-fracton  correlated events, for which the leading amplitude at $ g^2 + i 0^{\pm}$  are given by \cite{CDUlong}
\begin{align} 
\label{ff-ambiguity}
[\F_{i}  \bar \F_{i} ]_{\th=0^{\pm}} &= \Re\,[\F_{i}  \bar \F_{i} ] + i \,
   \Im\, [\F_{i}  \bar \F_{i} ]_{\th=0^{\pm}}  \\
&= \left[ \log  \left(   \frac{\lambda}{16 \pi } \right)  - \gamma  \right] 
\frac{16 }{\lambda}  e^{-\frac{16 \pi}{\l}} 
 \pm  i  \frac{32  \pi}{\lambda}   e^{-\frac{16 \pi}{\l}}  \nonumber
\end{align}
This leads to the cancellation of the non-perturbative ambiguities coming from perturbation theory against  the ambiguity that arises from the NP saddle.  That is: 
\begin{align} 
 \Im\,\B_{0, \th=0^\pm}  +   \Im\, [ \F_{i}  \bar \F_{i}   ]_{\th=0^{\pm}} = 0 .
 \label{confluence} 
\end{align} 
This is a QFT example of Borel-\'Ecalle resummation, a generalization of Borel resummation 
to account for the Stokes phenomenon.

{\bf Mass gap on $\R \times S^1$ and $\R^2$:}
A speculation by 't Hooft that  IR renormalons may be related to the mass gap and confinement in QCD \cite{tHooft:1977am} finds a concrete realization in our  approach.
The leading ambiguity on $\R^2$ is proportional to 
   $e^{ -  \frac{ 8\pi  }{g^2(Q) N} } \sim \Lambda^2/Q^2 $, and recent works  \cite{Argyres:2012vv,Argyres:2012ka, Dunne:2012ae} have shown that, in the semi-classical domain,  it is always  ``half" of the renormalon  which leads to a mass gap. 
   If we assume that this semi-classical fact extrapolates to the strongly coupled domain,   we observe that indeed,  $e^{ -  \frac{ 4\pi  }{g^2(Q) N} } \sim \Lambda/Q $, proportional to the first power of the strong scale.   
     In our current example, in the semi-classical domain, the mass gap is a one-fracton (half-renormalon) effect  and is given by $m_g \sim \frac{1}{LN}  e^{-\frac{8 \pi}{\l}} \sim  \Lambda (\Lambda LN)$ for $LN\Lambda \lesssim 2\pi$.  In future work, it would be important to understand fully the origin of the dilution factor highlighted in Eq.~\eqref{eq:s1} as the theory moves continuously from the semi-classical domain to the strongly coupled domain. 
       
{\bf Conclusions:}
Resurgence theory shows that in the principal chiral model (an asymptotically free theory without instantons) standard homotopy considerations are insufficient to classify saddle points in the path integral. Requiring the model to be well defined in the sense of Borel-\'Ecalle summability \cite{Dunne:2012ae,Dunne:2012zk,Argyres:2012ka,Argyres:2012vv},
together with the physical principle of continuity and spatial
compactification \cite{Unsal:2008ch}, leads to the existence of new
non-BPS `fracton' solutions  giving a semiclassical realization of IR
renormalons, and also provides  a quantum interpretation to the
classical  uniton solutions. The
 fracton contributions to the path integral of the PCM give the microscopic origin of the mass gap of the theory. 
 \\
 
{\it Acknowledgements.}
We acknowledge support from U.S. DOE grants 
 FG02-94ER40823 (A.C.),  DE-FG02-92ER40716 (G.D.),  DE-FG02-12ER41806 (M.\"U.) and
European Research Council Advanced Grant No. 247252 ``Properties and Applications of the Gauge/Gravity Correspondence'' (D.D.).

\bibliographystyle{apsrev4-1}
\bibliography{PCM} 

\end{document}